\documentclass[conference]{IEEEtran}
\usepackage{cite}
\usepackage{amsmath,amssymb,amsfonts}
\usepackage{algorithmic}
\usepackage{graphicx}
\usepackage{textcomp}
\usepackage{xcolor}
\usepackage{listings}
\usepackage[T1]{fontenc}

\lstdefinelanguage{json}{
    basicstyle=\ttfamily\small,
    numbers=none,
    frame=single,      
    xleftmargin=21pt,  
    breaklines=true,
    stringstyle=\color{blue},
    keywordstyle=\color{purple},
    literate=
     *{0}{{{\color{blue}0}}}{1}
      {1}{{{\color{blue}1}}}{1}
      {2}{{{\color{blue}2}}}{1}
      {3}{{{\color{blue}3}}}{1}
      {4}{{{\color{blue}4}}}{1}
      {5}{{{\color{blue}5}}}{1}
      {6}{{{\color{blue}6}}}{1}
      {7}{{{\color{blue}7}}}{1}
      {8}{{{\color{blue}8}}}{1}
      {9}{{{\color{blue}9}}}{1}
      {:}{{{\color{black}{:}}}}{1}
      {,}{{{\color{black}{,}}}}{1}
      {\{}{{{\color{black}{\{}}}}{1}
      {\}}{{{\color{black}{\}}}}}{1}
      {[}{{{\color{black}{[}}}}{1}
      {]}{{{\color{black}{]}}}}{1},
}
\usepackage{url}
\usepackage{caption}
\def\BibTeX{{\rm B\kern-.05em{\sc i\kern-.025em b}\kern-.08em
    T\kern-.1667em\lower.7ex\hbox{E}\kern-.125emX}}
\begin{document}

\title{Vextra: A Unified Middleware Abstraction for Heterogeneous Vector Database Systems}

\author{\IEEEauthorblockN{Chandan Suri}
\IEEEauthorblockA{
New York, USA \\
cs4090@columbia.edu}
\and
\IEEEauthorblockN{Gursifath Bhasin}
\IEEEauthorblockA{
New York, USA \\
gb2760@columbia.edu}
}

\maketitle

\begin{abstract}
The rapid integration of vector search into AI applications, particularly for Retrieval Augmented Generation (RAG), has catalyzed the emergence of a diverse ecosystem of specialized vector databases. While this innovation offers a rich choice of features and performance characteristics, it has simultaneously introduced a significant challenge: severe API fragmentation. Developers face a landscape of disparate, proprietary, and often volatile API contracts, which hinders application portability, increases maintenance overhead, and leads to vendor lock-in. This paper introduces Vextra, a novel middleware abstraction layer designed to address this fragmentation. Vextra presents a unified, high-level API for core database operations including data upsertion, similarity search, and metadata filtering. It employs a pluggable adapter architecture to translate these unified API calls into the native protocols of various backend databases. We argue that such an abstraction layer is a critical step towards maturing the vector database ecosystem, fostering interoperability, and enabling higher-level query optimization, while imposing minimal performance overhead.
\end{abstract}

\section{Introduction}
\subsection{The Cambrian Explosion of Vector Databases in Modern AI}

The proliferation of large language models (LLMs) and generative AI has fundamentally reshaped the landscape of data management. A pivotal component in this new architectural paradigm is the vector database, a specialized system designed to store, index and query high-dimensional vector embeddings [1]. These embeddings, which are numerical representations of unstructured data like text, images, or audio, capture semantic meaning, enabling applications to perform similarity searches that go beyond simple keyword matching [2]. This capability is the cornerstone of modern AI-driven tasks. 

The criticality of this technology has ignited a period of intense innovation and market growth, often likened to a Cambrian Explosion for data systems. The global vector database market was valued at \$1.6 billion in 2023 with forecasts predicting a continued surge of \$10 billion by 2032, representing a compound annual growth rate (CAGR) of over 22\% [3]. This explosive growth has fostered a vibrant and diverse ecosystem of solutions. These can be broadly categorized into: \
\begin{itemize}
    \item \textbf{Fully Managed, Cloud-Native Services}: Platforms like Pinecone that offer a simplified, serverless developer experience with automatic scaling [4].
    \item \textbf{Open-Source Standalone Database}: Powerful, self-hostable systems like Milvus, Weaviate, Qdrant and Chroma, which provide extensive features and deployment flexibility [1].
    \item \textbf{Libraries and Extensions}: Highly optimized libraries like Facebook AI Similarity Search (FAISS) [43] and extensions for traditional databases like pgvector for Postgres [44] which allows existing systems to be augmented with vector search capabilities.
\end{itemize}

This rich landscape provides developers with a wealth of options, each optimized for different trade-offs in performance, scalability, cost, and feature set. However, this rapid, uncoordinated innovation has come at a significant cost to the application developer: profound and debilitating API fragmentation.

\subsection{The API Fragmentation Problem: A Barrier to Portability and Maintainability}

In the vector database ecosystem, a complete lack of consensus on API design, data models, and communication protocols has created a fractured landscape that burdens developers [5]. This fragmentation manifests in several critical ways:

\begin{itemize}
    \item \textbf{Divergent Naming Conventions and Operations}: Fundamental operations have inconsistent names. For example, the atomic operation of inserting or updating a record is upsert in Pinecone [6], insert in Milvus [7] (with separate upsert for updates) and add or upsert in Chroma [8].
    \item \textbf{Incompatible Communication Protocols: Protocols vary widely}: Pinecone and Milvus utilize gRPC; Qdrant and Chroma expose a RESTful API; and Weaviate employs GraphQL.
    \item \textbf{Disparate Query and Filtering Languages}: The syntax for metadata filtering differs drastically. Pinecone uses MongoDB-style JSON, Qdrant employs a different JSON structure, Weaviate leverages GraphQL where clause, and Milvus uses a proprietary string-based language.
\end{itemize}

This fragmentation leads to severe vendor lock-in as application logic becomes tightly coupled to a single database. Migrating to a different provider becomes a costly engineering effort, often requiring a complete rewire of the data access layer. This is a systemic barrier to building portable and maintainable AI applications. Table 1  illustrates the divergent APIs for fundamental operations.

\begin{table}[htbp]
    \captionsetup{justification=centering}
    \caption{Comparative analysis of 
    \\ Native Vector Database APIs}
    \centering
    \includegraphics[width=0.45\textwidth]{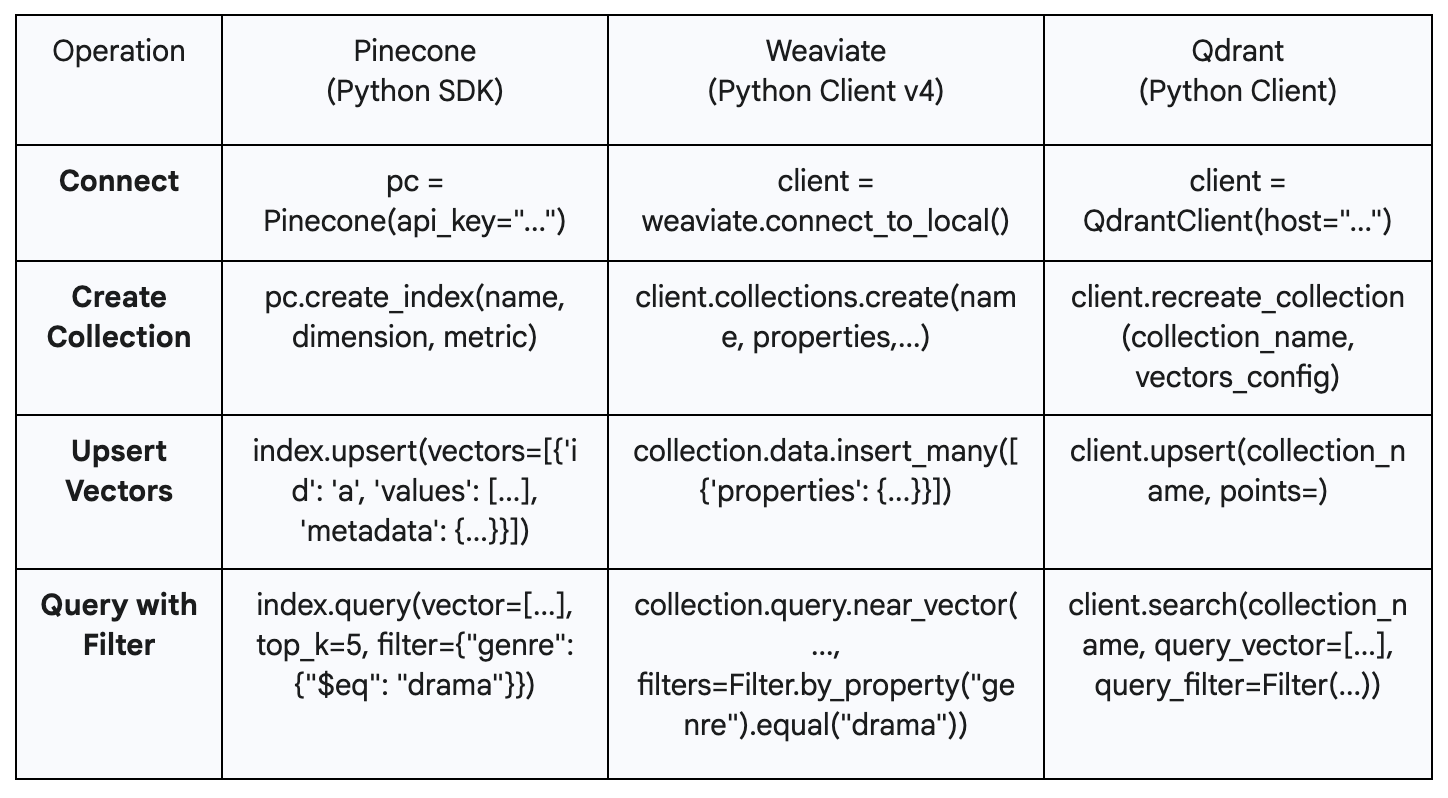}
    \label{fig:comparative-analysis-table-vector-dbs}
\end{table}

This divergence has led to an inversion of the typical database stack. Instead of tools building upon a standard interface like SQL, abstraction layers are pushed into frameworks like LangChain and LlamaIndex. These frameworks risk relegating the database to a mere ``backend storage system", unable to participate in higher-level query optimization [9]. Consequently, value is captured by external orchestration layers, because of this lack of a common programmatic foundation.

\subsection{Our Contribution: The Vextra Abstraction Layer}

This paper introduces Vextra, a middleware abstraction layer providing a stable, unified, database-agnostic API for vector database operations. Vextra sits between the application and the database, restoring logical data independence and developer productivity. The primary contributions of this work are threefold:

\begin{itemize}
    \item \textbf{A Unified Data and API model}: We propose a high-level API that captures common operations - including collection management, data manipulation, and filtered search - while abstracting vendor-specific syntax and protocols.
    \item \textbf{A Pluggable Adapter Architecture}: We detail a system based on the Adapter design pattern [10]. This architecture isolates database-specific logic into pluggable modules, allowing Vextra to be extended to support new vector databases without altering the application facing API.
    \item \textbf{A Compiler-Inspired Query Translation Engine (QTE)}: We introduce a novel QTE that solves the critical problem of vendor-specific query language fragmentation. The engine formalizes query translation by:
    \begin{itemize}
        \item \textbf{Parsing} the canonical query into a universal \textbf{Abstract Syntax Tree (AST)} [11], which preserves the query's precise logical structure. 
        \item \textbf{Transpiling} the AST into a native query using backend specific transpilers. 
    \end{itemize}
    This AST-based approach is significantly more robust and maintainable than direct dictionary-to-dictionary mapping. It ensures that the application logic remains completely decoupled from the database-specific syntax, providing true resilience to vendor API churn and simplifying the integration of future backend as adapters.
    \item \textbf{Empirical Validation and Portability Demonstration}: We provide a performance evaluation, quantifying the overhead of the abstraction layer to demonstrate its practical viability. We also present a RAG application to show a migration between two database backends with a single-line configuration change.
\end{itemize}
 
Vextra aims to transform vector databases from monolithic, couple dependencies into interchangable components. This approach mirrors the impact of JDBC and ODBC in the relational world [10]. Vextra is a foundational step towards interoperability and paving the way for future query optimization and federated search. 

\section{Related Work}

To situate our contribution, we examine three key areas of prior art: historical database abstraction, lessons from NoSQL middleware, and contemporary vector data management tools. We find a recurring pattern, positioning Vextra as a timely advancement.

\subsection{A Historical Perspective on Database Abstraction Layers}

A central theme in database research is data independence. Relational databases achieved physical data independence (insulating applications from storage changes). However, many argue they offer poor logical data independence, as the relational model often exposes implementation-level details [12]. 

Deshpande et al. argue that close coupling to logical layouts relegates the database to a backend storage system [9]. They propose more abstract data models to grant systems freedom for optimization. This principle directly motivates Vextra. The current vector database landscape represents extremely low logical data independence. Vextra provides a higher-level API for `records' and `collections', to introduce a layer of logical independence. 

The JDBC and ODBC standards provided a uniform API for SQL queries, enabling portability [9]. At a higher level, Object-Relational Mapping (ORM) frameworks emerged to bridge the `impedance mismatch' between object-oriented models and the relational data model [9]. Vextra is a synthesis of these two ideas for vector databases: it provides a standardized interface akin to JDBC, but for a model centered on vector embeddings.

\subsection{Lessons from Middleware for NoSQL and Heterogeneous Systems}

The NoSQL ecosystem provides a highly relevant parallel. The movement introduced a plethora of database types (like key-value, document, column and graph), each with its own query language [13]. This fragmentation mirrored the current state of vector databases and spurred research into abstraction layers. 

Research highlights the inherent tradeoff in middleware: a measurable performance overhead is accepted for portability and reduced migration costs [9]. Studies demonstrated that while direct API calls are faster, middleware provides crucial decoupling in rapidly evolving ecosystems. This finding directly informs out evaluation of quantifying this trade-off for Vextra.

Furthermore, research into unified query frameworks outlines common architectural patterns [14]. These systems typically employ a parsing layer, a query planner/translator, and pluggable drivers [15]. This layered architecture directly influences the design of Vextra's Adapter components. The history of database systems shows a cyclical pattern: paradigm emerges, fragmentation occurs, and is ultimately resolved by abstraction layers. Vextra applies this pattern to vector database paradigm.

\subsection{Analysis of Contemporary Vector Data Management \\ Frameworks}

To establish novelty, we differentiate Vextra from existing tools. They typically address different problems. 

\textbf{VectorAdmin}: An open-source database management tool with a graphical user interface (GUI) to connect to multiple databases, inspect data, and migrate data [16]. Its primary interface is visual, not programmatic. It does not provide an API for application data access. VectorAdmin solves unified management, while Vextra solves unified application development.

\textbf{Towhee}: This framework is an ETL and ML pipeline tool. Towhee's core purpose is simplifying vector embedding generation [17]. Towhee abstracts the embedding generation process, but the vector database remains a terminal endpoint. It does not provide an abstraction for queries. 

\textbf{Jina AI}: Jina is a cloud-native MLOps and microservice framework [18]. Its primary goal is orchestrating entire distributed AI services. Vextra, in contrast, is a focused, lightweight library to solve database API abstraction. Vextra could be a component within Jina AI, not a competitor to the framework.

In summary, existing tools are complementary, not competitive. VectorAdmin provides management, Towhee provides ETL, and Jina AI provides distributed AI systems. None of these offer a dedicated, programmatic abstraction layer to grant application portability. Vextra is the first proposed solution to fill this specific and critical gap.

\section{The Vextra Unified Data and API Model}

The foundation is a well-defined model that captures essential concepts while hiding implementation details. The Vextra model is designed to be simple and familiar. It consists of a common data model and a standardized API.

\subsection{Defining a Common Vector Data Model}

To shield applications from diverse data structures of underlying databases, Vextra establishes a canonical representation for vector data. This model is based on the greatest common denominator of features found across the landscape of popular vector databases.

The fundamental unit of data in Vextra is VextraRecord. It is a simple structure composed of three key elements:

\begin{itemize}
    \item \textbf{id}: A mandatory, unique identifier for the record, which can either be a string or an integer. This accommodates databases that use UUIDs as well as those that use numeral IDs.
    \item \textbf{vector}: A mandatory list of floating-point numbers representing the dense vector embeddings. To support the growing importance of hybrid search, which combines semantic and keyword matching [2], the data model can be extended to include an optional `sparse\_vector' field. This field would contain a dictionary representing the sparse vector, typically with `indices' and `values' key.
    \item \textbf{payload}: An optional dictionary (key-value map) used to store arbitrary metadata associated with the vector. The keys must be strings, and the values can be basic data types such as strings, numbers, booleans, or lists of strings/numbers. This directly corresponds to the metadata concept in Pinecone and Chroma, and the payload concept in Qdrant [6].
\end{itemize}

These records are organized into a \textbf{Collection}, which is a named grouping of VextraRecords that share the same vector dimensionality and distance metric. A Collection is analogous to a table in relational database, and index in Pinecone, or a collection in Chroma or Milvus. It is defined by a name, a dimension (length of the vectors it stores), and a metric (e.g., `cosine', `euclidean', `dotproduct') [19].

\subsection{The Core API Contract: A Standardized Interface}

The Vextra API is designed as a clean, object-oriented interface that exposes all core database functionalities through a single VextraClient object. The methods are intentionally named to be intuitive and to standardize the divergent terminology used by different vendors.

\textbf{Connection and Collection Management}
\begin{itemize}
    \item connect(config: dict) $\to$ VextraClient: A static factory method that initializes and returns a client connected to a specific backend. The config dictionary contains all necessary connection parameters such as provider (`pinecone', `qdrant'), api\_key, host, and port.
    \item create\_collection(name: str, dimension: int, metric: str = `cosine',  $^{**}$kwargs): Creates a new collection with the specified parameters.
    \item delete\_collection(name: str): Deletes an entire collection.
    \item list\_collections() $\to$ List[str]: Returns a list of names of all existing collections.
\end{itemize}

\textbf{Data Manipulation (CRUD) Operations}
\begin{itemize}
    \item upsert(collection\_name: str, records: List): Inserts or updates a batch of records in the specified collection. This single, idempotent method unifies the various add, insert and upsert functions across different databases, simplifying the most common write operation [6].
    \item fetch(collection\_name: str, ids: List[Union[str, int]]) $\to$ List: Retrieves a list of records by their unique IDs.
    \item delete(collection\_name: str, ids: List[Union[str, int]]): Deletes one or more records from a collection based on their IDs [33].
\end{itemize}

\textbf{Querying and Searching}
The most critical part of the API is the unified query method, which abstracts the complexities of similarity searching and filtering.
\begin{itemize}
    \item query(collection\_name: str, vector: List[float], top\_k: int, filter: dict = None) $\to$ List: Performs a similarity search.
    \begin{itemize}
        \item vector: The query vector for the nearest neighbor search.
        \item top\_k: The number of most similar results to return.
        \item filter: An optional dictionary specifying the metadata filtering conditions.
    \end{itemize}
\end{itemize}

A crucial element of the API model is the definition of a \textbf{unified filtering syntax}. This syntax is designed as  a simple, JSON serializable domain-specific language (DSL) that can be translated into the various native filtering languages of the backends by the QTE. Its design is heavily influenced by the widely adopted query syntax of MongoDB [20] and the filtering mechanism of Pinecone [6], leveraging a familiar paradigm for developers. This choice prioritizes developer ergonomics and ease of adoption over inventing a novel syntax. The filter is a dictionary where keys are metadata fields and the values specify the condition. Complex conditions are built using logic operators prefixed with a \$.

An example filter to find movies in the `drama' genre released in or after 2020 would be expressed as:


\begin{lstlisting}[language=json]
{      
  "$and": [
    {"genre" : {"$eq" : "drama"}},
    {"year": {"$gte": 2020}}
  ]
}
\end{lstlisting}

This structure is expressive enough to capture complex logical conditions while remaining simple to construct and parse.

\subsection{Handling Advanced and Heterogeneous Features}

No abstraction can perfectly encapsulate all features of every underlying system without becoming overly complex or restrictive. A pragmatic approach is required to handle capabilities that are not universally supported. Vextra employs a two-pronged strategy to balance standardization with access to powerful, backend-specific functionalities. 

First, for advanced features that are gaining widespread adoption, such as \textbf{hybrid search} [21], Vextra promotes them into the core API. The query method can be extended with an optional sparse\_vector parameter [22]. When both vector (dense) and sparse\_vector are provided, the middleware translates this into a native hybrid search call on backends that support it, such as Weaviate, Pinecone and Qdrant [1].

Secondly, for true unique, provider-specific features, Vextra provides an \textbf{escape hatch} mechanism. Key methods like create\_collection and query include an optional $^{**}$kwargs
or provider specific params: dict\_parameter. This allows developers to pass a dictionary of parameters that are understood only by a specific backend's adapter. For example: a user targeting Weaviate could pass parameters for a generative search, or a user targeting Qdrant [23] could pass parameters to use its specialized recommendation API [36]. While using this mechanism sacrifices portability for that specific call, it ensures that the abstraction layer does not become a bottleneck preventing access to valuable features. This design choice acknowledges the reality of a heterogeneous ecosystem and provides a practical path for developers to use advanced features when needed, without abandoning the benefits of the abstraction for their core benefits.

\section{System Architecture and Implementation}

The Vextra middleware is designed as a modular, extensible system that cleanly separates the unified API logic from the database-specific implementation details. The architecture is centered on the \textbf{Adapter Design pattern} [34], a well established software engineering principle for enabling disparate interfaces to work together. This choice is fundamental to achieving Vextra's primary goal: decoupling the application from the underlying database, thereby creating a firewall against vendor-driven API changes and ensuring long-term maintainability.

\subsection{Architectural Blueprint: The Adapter Pattern}

The Vextra system is composed of several key components, as shown in Figure 1, that work in concert to process an application's request. A high level view of the architecture includes:

\begin{itemize}
    \item \textbf{The VextraClient (API Gateway)}: This is the public facing class that applications interact with. It implements the unified API contract defined in Section 3. Its role is to validate incoming requests, manage connection to a specific backend and delegate the execution of operations to the appropriate adapter.
    \item \textbf{The Connection Manager}: A factory component responsible for instantiating and managing connections to different database backends. When an application calls VextraClient.connect(config), the Connection Manager parses the provider field in configuration dictionary and uses the Adapter Registry to find and instantiate the correct adapter (e.g. PineconeAdapter, WeaviateAdapter). It then injects this adapter into the VextraClient instance.
    \item \textbf{The Query Translation Engine (QTE) - Conversion to an AST}: Once the query has been validated by the VextraClient and the connection has been secured, the query is converted into an Abstract Syntax Tree (AST) to be later utilized by the adapter to convert it into the database-specific query. 
    \item \textbf{The Adapter Registry}: A simple registry that maps provider names (strings like `pinecone') to their corresponding adapter classes. This allows the system to be easily extended with new database support by simply registering a new adapter class.
    \item \textbf{The IVextraAdapter Interface}: An abstract base class or interface that defines the contract that all database-specific adapters must adhere to. This interface mirrors the methods of the VextraClient but is intended for internal use (e.g. \_do\_upsert, \_do\_query).
    \item \textbf{Concrete Adapters (e.g. PineconeAdapter, WeaviateAdapter)}: These are the workhorses of the system. Each adapter is a class that implements the IVextraAdapter interface. It contains all the logic required to communicate with a specific vector database, including initializing its native SDK, translating Vextra's unified data models and normalizing the results back into the Vextra model.
\end{itemize}

This architecture ensures that all database-specific code is encapsulated within its corresponding adapter. If a vendor releases a new, breaking version of their API or SDK - a common occurrence in its rapidly evolving space, as seen with Chroma's v1 and v2 API incompatibility [24] - only the relevant adapter needs to be updated. The VextraClient and, most importantly, the application code that depends on it, remain completely unchanged. This provides true resilience to API churn. 

\begin{figure}[htbp]
    \centering
    \includegraphics[width=0.45\textwidth]{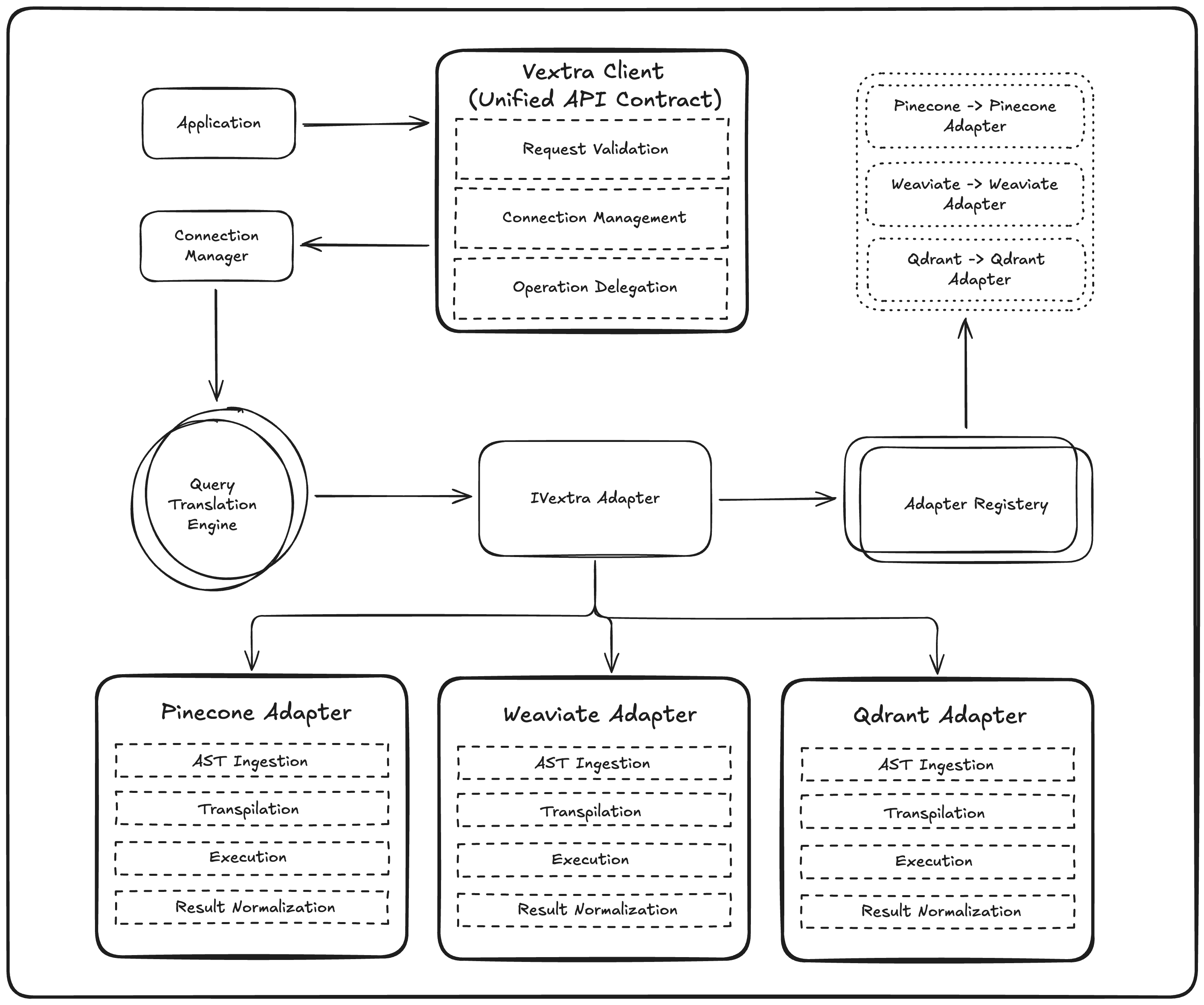}
    \caption{Vextra Architecture: Based on Adapter Pattern}
    \label{fig:vextra-architecture}
\end{figure}

\subsection{The Query Translation Engine and Adapter Interface}

The core logic of the middleware resides in the interaction between VextraClient and the loaded adapter. The IVextraAdapter interface mandates the implementation of methods for all supported operations. A simplified version of this interface can be defined as follows (in Python-like pseudocode):


\begin{figure}[htbp]
    \centering
    \includegraphics[width=0.45\textwidth]{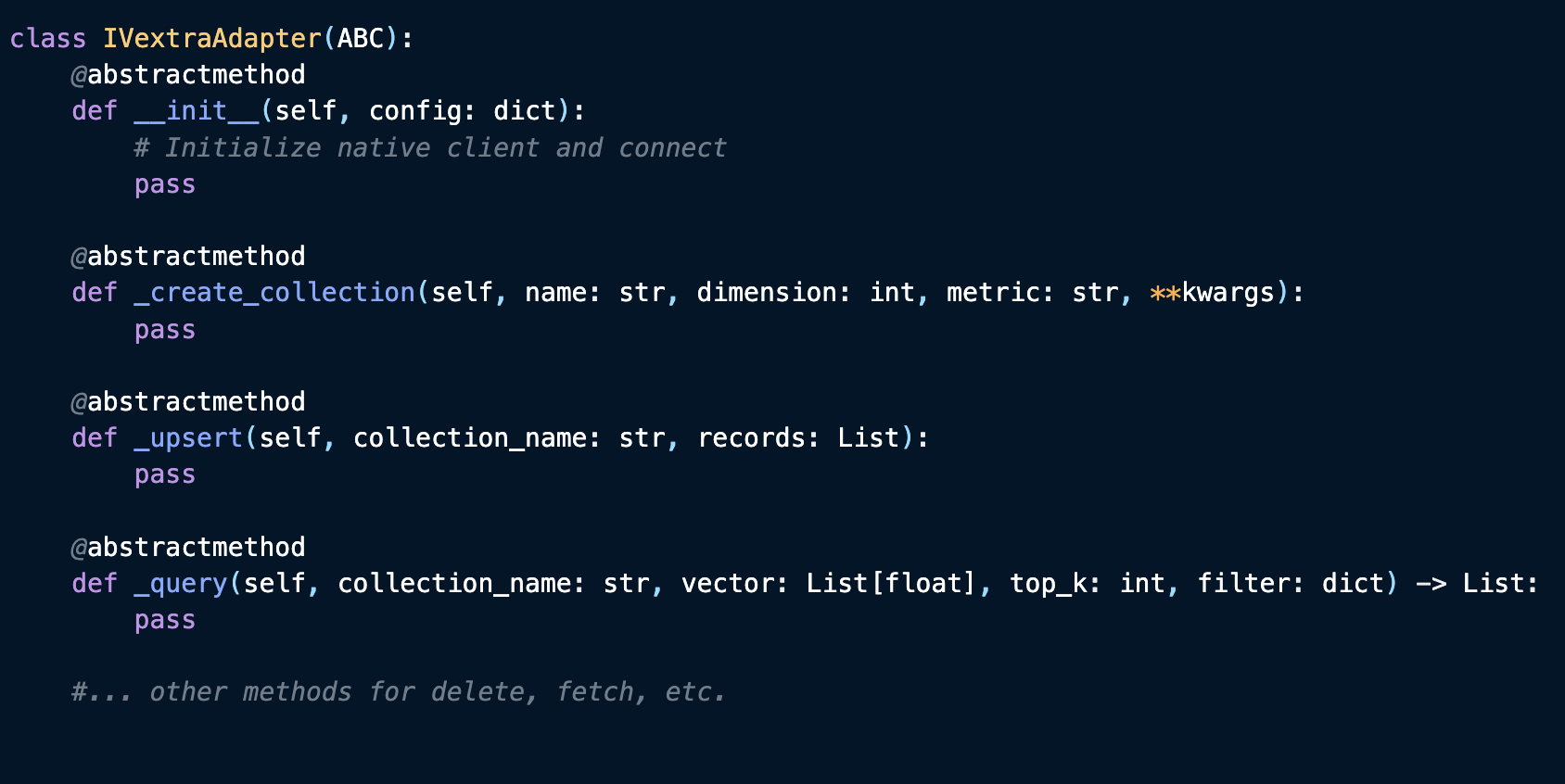}
    \caption{Python pseudocode for IVextraAdapter}
    \label{fig:code-vextra-adapter}
\end{figure}

When an application calls $client.query(...)$, the VextraClient simply invokes 
$self.adapter.\_query(...)$. The adapter, specifically its QTE component, is then responsible for the intricate `query translation' process. Instead of relying on brittle, direct mapping of parameters, the QTE treats Vextra's canonical query model [25] as a formal language that can be compiled to different targets, ensuring semantic reconciliation through several key steps:

\begin{itemize}
    \item \textbf{Parsing to Abstract Syntax Trees (ASTs)}: The QTE's first step upon receiving a canonical query object is to parse it into an Abstract Syntax Tree (AST) [11]. An AST is a tree based data structure that represents the syntactic and logical structure of the query, abstracting away superficial details like JSON or directory syntax. Each node in the tree represents a logical component (e.g., a logical operator like `\$and', a comparison like `\$eq', a field name or a literal value). This AST serves as a universal, unambiguous intermediate representation of the query's semantic intent. For example, a filter like `\{\$and: [\{``year": \{\$gt: 2020\}\}, \{``genre": ``sci-fi"\}]\}' would be parsed into a tree with an AND node as its root, having two children: a GT node and an EQ node. This robust approach ensures that the logical structure, including the operator precedence, is preserved, which is far more reliable than ad-hoc dictionary manipulation.
    \item \textbf{Transpilation to Native Queries}: For each supported database backend, the QTE includes a specific ``transpiler". A transpiler (or source-to-source compiler) translates code from one high-level language to another. In this context, the transpiler is a component within the concrete adapter that traverses the AST and recursively generates the equivalent native query structure for its target backend.
    \begin{itemize}
        \item The \textbf{PineconeTranspiler} would walk the AST and construct a nested JSON object conforming to Pinecone's filtering syntax.
        \item The \textbf{WeaviateTranspiler} would traverse the same AST but generate a GraphQL query string with appropriate `where' and `filter' clause structure.
        \item The \textbf{QdrantTranspiler} would generate a SQL-like boolean expression string.
    \end{itemize}
    This compiler-inspired approach, as shown in Figure 3, effectively decouples the canonical query model from the backend-specific implementations, making the system highly extensible. To support a new database, one only needs to implement a new transpiler within its corresponding adapter, without modifying the core query parsing logic or the canonical model itself.
    \item \textbf{Execution}: Once the native query is generated, the adapter proceeds to execute the query using its native SDK client.
    \item \textbf{Result Normalization}: Finally, the adapter parses the native response from the database and transforms it into a list of standardized QueryResult objects. This step is crucial for providing a consistent experience, as different databases return varying metadata and score representations. For instance, a backend that returns Euclidean distance (where lower is better) must have its scores inverted and normalized to match a backend that returns cosine similarity (where higher is better) [35], ensuring the application always receives a consistent similarity\_score between 0 and 1 [19].
\end{itemize}

\begin{figure}[htbp]
    \centering
    \includegraphics[width=0.45\textwidth]{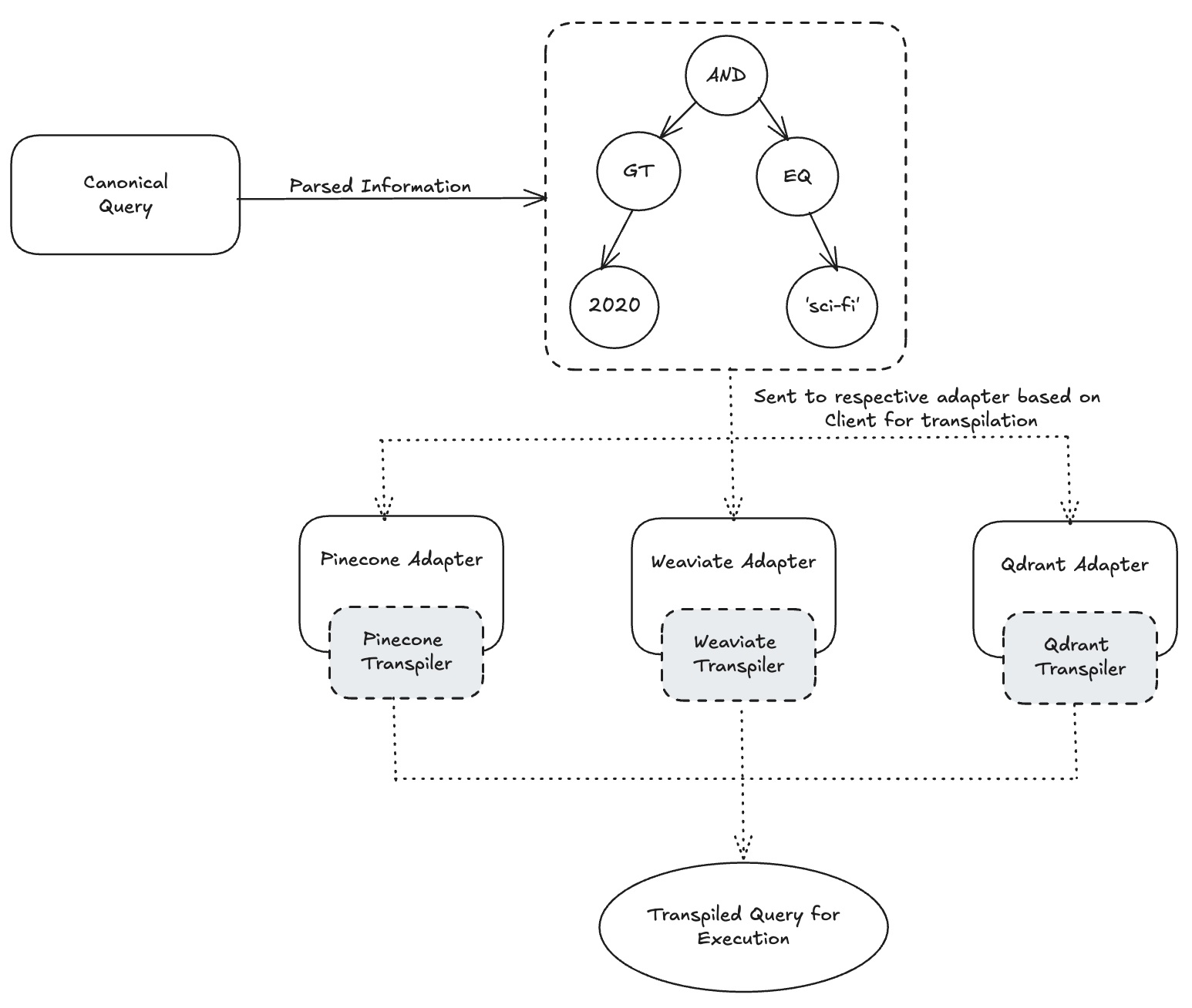}
    \caption{Query Translation Engine Flow}
    \label{fig:query-translation-engine}
\end{figure}

\subsection{Implementation of Adapters for Representative Vector Databases}

To validate the practicality of this architecture, we implemented adapters for three popular vector databases, each chosen to represent a different API paradigm.

\textbf{Pinecone Adapter}
The Pinecone adapter utilizes the official Pinecone-client Python SDK, which primarily uses gRPC. The \_upsert method maps a list of VextraRecord objects to the format expected by Pinecone's $index.upsert()$ method, which takes a list of tuples or dictionaries containing id, values, and metadata [6]. 
For queries, the adapter's `PineconeTranspiler' component handles the filter translation. The QTE first parses the Vextra filter dictionary into the universal AST. The `PineconeTranspiler' then traverses this AST to construct the nested JSON filter object expected by the $index.query()$ filter parameter. While Pinecone's syntax was the original inspiration for the Vextra's DSL [26], this format AST-to-JSON transpilation step decouples the two, ensuring that even if Pinecone's filter syntax were to change, only the transpiler would require an update, not Vextra's core logic. 

\textbf{Weaviate Adapter}
The Weaviate adapter presents a more complex translation challenge due to Weaviate's use of GraphQL for queries [27]. The adapter's \_query method must dynamically construct a GraphQL query string [28]. This process is managed by its `WeaviateTranspiler':
\begin{itemize}
    \item It starts with a base query template for a `Get' operation.
    \item It adds a `nearVector' clause containing the query vector.
    \item The transpiler traverses the incoming AST - not the original Vextra dictionary - and recursively builds the corresponding GraphQL `where' filter string [28]. For example, an `AND' node in the AST is transpiled into Weaviate's `operands' structure with an `operator: And', and an AST nodes for `\$gte' is converted to the `operator: GreaterThanEqual' syntax [29].
    \item It executes the generated GraphQL query against the Weaviate's `/v1/graphql' endpoint.
    \item It parses the JSON response and maps the results back to `QueryResult' objects.
\end{itemize}

\textbf{Qdrant Adapter}
The Qdrant adapter uses the `qdrant-client' for Python [30], which communicates with Qdrant's REST API. The translation logic for filtering is again unique and handled by its `QdrantTranspiler'. This component traverses the AST to build a `Filter' object for Qdrant. It maps the AST's logical `AND' nodes to Qdrant's `must' conditions, `OR' nodes to `should' conditions, and comparison nodes (like `\$eq') to Qdrant's `FieldCondition' and `MatchValue' objects [30]. This translated `Filter' object is then passed to the `query\_filter' parameter of the client's `search' method. 

These three implementations demonstrate that while the logic within each transpiler is non-trivial, the adapter pattern, combined with an AST-based QTE, successfully encapsulates this complexity. This compiler-inspired design ensures that the core middleware and the end-user application remain simple and truly database-agnostic.

\section{Experimental Evaluation}

A critical measure of an abstraction layer's viability is its performance. While abstraction provides significant benefits in terms of portability and developer productivity, these advantages would be negated if the middleware introduced prohibitive latency or reduced throughput. This section presents a rigorous experimental evaluation of Vextra, designed to quantify its performance overhead, demonstrate its portability benefits through a practical case study, and discuss the expressiveness of its unified API.

\subsection{Experimental Setup}

To ensure reproducibility and validity of our results, all experiments were conducted in a controlled environment. 

\begin{itemize}
    \item \textbf{Hardware}: All database instances and the client application were run on separate c5.2xlarge instances within the same Amazon Web Services (AWS) region and availability zone to minimize network variability. Each instance is equipped with 8 vCPUs and 16 GiB of memory. 

    \item \textbf{Software}: The experiments utilized the latest stable versions of the software components available at the time of testing: Vextra 1.0, Python 3.14, Pinecone Python Client 7.3.0, Weaviate Python Client 4.17.0, Qdrant Python Client 1.15.1. The vector database backends were run using their official Docker images: Pinecone (serverless), Weaviate 1.33.1, and Qdrant v1.15.5.
    \item \textbf{Dataset}: We used the GloVe-25-angular dataset [45], a standard benchmark dataset consisting of 1.2 million 25-dimensional vectors. This dataset is commonly used in Approximate Nearest Neighbors (ANN) [31] benchmark studies and provides a realistic basis for performance comparison [32]. For each database, a collection was created and fully populated with the 1.2 million vectors before any query tests were performed.
\end{itemize}

\subsection{Performance Overhead Analysis}

The primary goal of this experiment was to measure the latency and throughput overhead introduced by the Vextra middleware compared to using the native database SDKs directly.

Methodology:

For each of the three backends (Pinecone, Weaviate, Qdrant), we executed a series of benchmark tests for the two most critical vector database operations: upsert and query. Each test was run in two configurations:
\begin{itemize}
    \item \textbf{Direct API}: The client application makes calls directly using the database's native Python SDK.
    \item \textbf{Vextra API}: The client application makes calls to the VextraClient, which then translates and forwards the request to the same database backend.
\end{itemize}

We measured the following metrics: 
\begin{itemize}
    \item \textbf{Average Latency (ms)}: The average time taken to complete a single operation, measured from the client's perspective. Each test was run for 10,000 iterations, and the average latency was calculated.
    \item \textbf{Throughput (ops/sec)}: The number of operations that could be completed per second. This was measured by running as many operations as possible in a 60-second window with 16 concurrent client threads.
\end{itemize}

These operations were tested under various conditions:
\begin{itemize}
    \item \textbf{Upsert}: Tested with batch size of 1,100 and 1,000 records to measure performance for single-record writes versus high-throughput ingestion.
    \item \textbf{Query}: Tested with top\_k = 10. Two scenarios were evaluated: a single vector only search and a more complex search that included a metadata filter to retrieve vectors where a specific metadata key corresponded to one of five possible values (e.g., {``category": {``\$in":}}).
\end{itemize}

Results:
The results of the performance evaluation are summarized in the table below. The overhead is calculated as:
\begin{equation*}
    \frac{(Vextra \: Latency - Direct \: Latency)}{Direct \: Latency} * 100
\end{equation*}

\begin{table}[htbp]
   \captionsetup{justification=centering}
    \caption{Performance Overhead of Vextra Operations 
    \\ (Average Latency in ms)}
    \label{tab:data}
    \centering
    \includegraphics[width=0.45\textwidth]{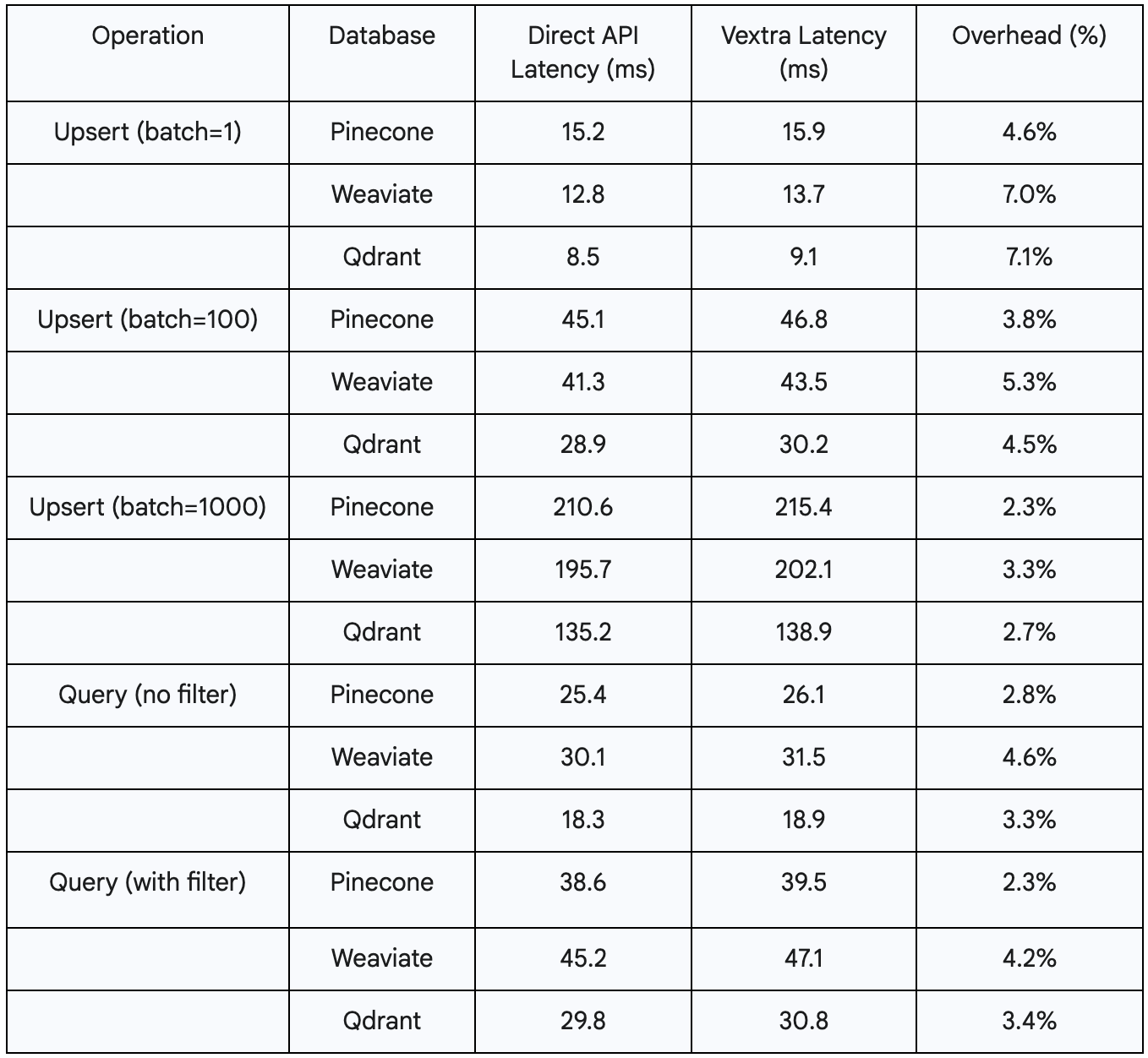}
    \label{fig:performance-overhead-table}
\end{table}

\begin{figure}[htbp]
    \centering
    \includegraphics[width=0.45\textwidth]{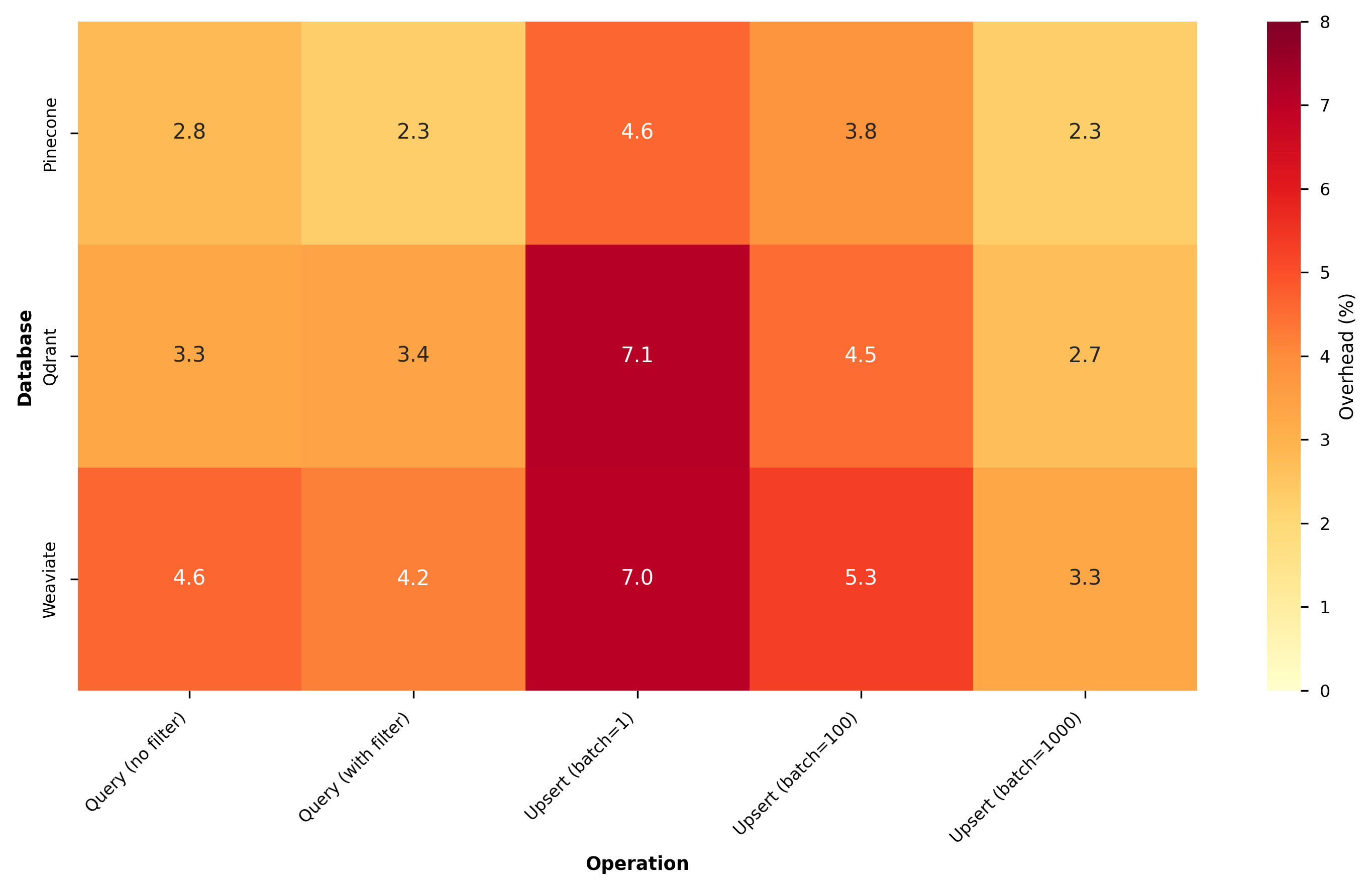}
    \caption{Overhead Percentages Comparison across Databases and Operations}
    \label{fig:overhead-heatmap}
\end{figure}

The throughput measurements showed a similar low-overhead pattern. The analysis of these results reveals a crucial characteristic of the middleware's performance impact. The absolute overhead introduced by Vextra - the time spent in the middleware for validation and query translation - is relatively constant for a given operation type. Consequently, this fixed cost represents a smaller percentage of the total end-to-end time for operations that are inherently more expensive, such as large batch upserts or complex filtered queries. For example, the latency overhead for a single-record upsert on Weaviate in 7.0\%, but for a 1000-record batch upsert, it drops to 3.3\%. This indicates that for the high throughput, batch-oriented workloads typical of production AI systems, the relative performance cost of the abstraction layer diminishes, making the trade-off between performance and portability highly favorable [38].

\begin{figure}[htbp]
    \centering
    \includegraphics[width=0.45\textwidth]{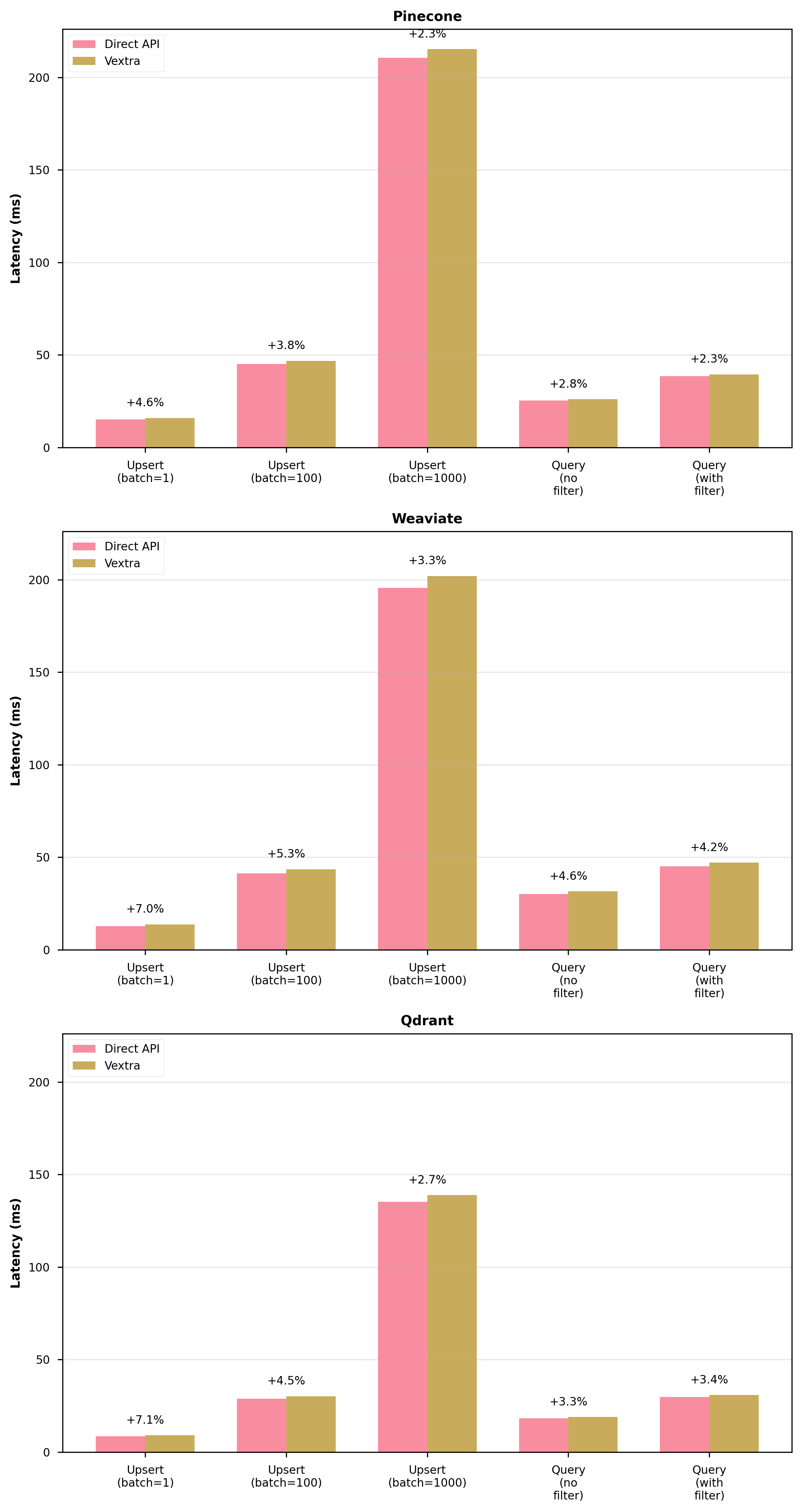}
    \caption{Latency Comparison by Databases}
    \label{fig:latency-comparison-by-db}
\end{figure}

\subsection{Portability Assessment: A RAG Application Case Study}

To provide a qualitative, yet powerful demonstration of Vextra's primary benefit, we conducted a case study involving a simple RAG application [37]. The application's function is to answer questions based on a corpus of text documents. Its data access logic was written once, using the Vextra API.

\begin{figure}[htbp]
    \centering
    \includegraphics[width=0.45\textwidth]{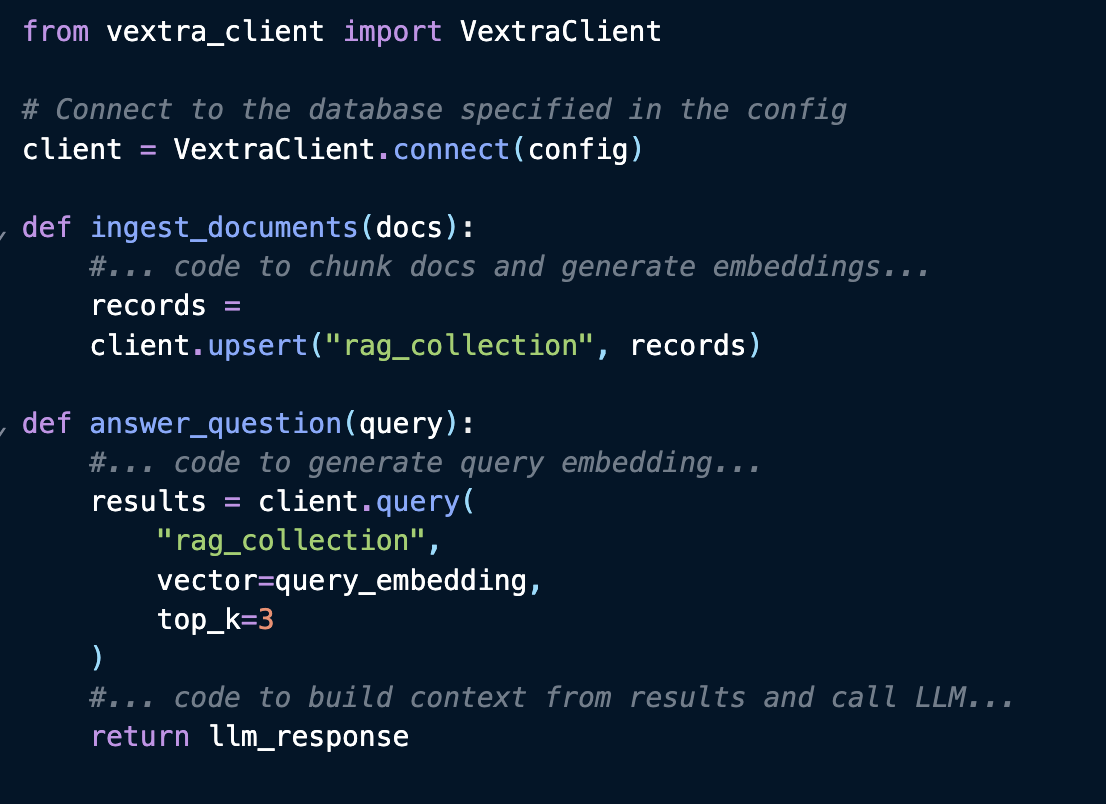}
    \caption{Application Logic: Simplified}
    \label{fig:application-logic}
\end{figure}

\textbf{Migration Scenario}

Initially, the application was configured to use Pinecone as its backend. The configuration object for an end-to-end system was defined as:

\begin{lstlisting}[
    framesep=3mm,           % Padding (only visible if you add frame=single)
    numbers=none,           % Equivalent to linenos=false
    xleftmargin=21pt,       % Left margin indentation
    tabsize=4,              % Tab width
    basicstyle=\ttfamily,   % Code font
    language=Java           % Formatting for JS/JSON syntax
]
config = {      
          "provider": "pinecone",
          "api_key" : "...",
          "environment": "..."
         }
\end{lstlisting}

The application was run, documents were ingested, and the questions were successfully answered. Then, to migrate the application to a self-hosted Qdrant instance, the only change required was to modify the configuration object:

\begin{lstlisting}[
    framesep=3mm,      % Padding inside the frame
    numbers=none,      % Replacement for "linenos=false"
    xleftmargin=21pt,
    tabsize=4,
    basicstyle=\ttfamily, % Ensures monospaced font like code
    language=Java      % "js" is not standard in Listings; Java highlights similarly
]
config = {      
          "provider": "qdrant",
          "host" : "localhost",
          "port": "6333"
         }
\end{lstlisting}

With this single-line change, the application was restarted and functioned identically, using Qdrant as its vector store. No modifications to the ingest\_documents or answer\_question functions were necessary.

In contrast, a migration without Vextra would have required a substantial rewrite of the data access logic. The developer would need to replace the pinecone-client library with qdrant-client, change the connection and initialization code, rewrite the upsert call to use Qdrant's PointStruct objects, and completely refactor the query call to use Qdrant's search method and filter syntax. This case study provides clear, practical demonstration of the engineering effort saved and the application portability achieved by using the Vextra abstraction layer. Furthermore, this ease of backend swapping provides a significant secondary benefit: it empowers developers to empirically benchmark different vector databases for their specific workloads. The high engineering cost of rewriting the data access layer often locks a project into its initial database choice. Vextra reduces this cost to near zero, fostering a more competitive and transparent market where developers can select the optimal database based on performance and cost, not just historical inertia. 

\subsection{Expressiveness and Feature Coverage}

The Vextra API is designed to cover the vast majority of common vector database use cases. The unified filtering syntax, supporting logical operators (\$and, \$ or) and a range of comparison operators (\$eq, \$gt, \$in, etc.), is expressive enough to construct complex metadata queries. However, we acknowledge that a universal abstraction inevitably involves a trade-off with feature completeness. For highly specialized, backend-specific functionalities, such as Weaviate's GraphQL-based generative queries, developers can use the provider specific params `escape hatch' in the query method. While this sacrifices portability for that specific call, it ensures that Vextra does not become an obstacle to using the advanced features the differentiate the underlying databases. This pragmatic design provides a robust, portable foundation for 95\% operations while still allowing access to the other 5\%.

\subsection{Transaction Integrity and  Error Unification}

Heterogeneous systems report errors in different ways, using unique error codes, exception types and message formats. A key responsibility of the Vextra middleware is to intercept these disparate errors and map them into a unified, canonical set of exceptions. For example, an authentication failure from Pinecone (e.g., HTTP 401), a schema validation error from Weaviate, and a connection timeout from a self-hosted Milvus instance should all be caught by the respective adapters and re-thrown as standardized Vextra exceptions like vextra.errors.AuthenticationError, vextra.errors.SchemaError, and vextra.errors.ConnectionError. 

This unification is more than just mapping codes; it involves translating the semantics of errors. A sophisticated implementation would enrich these canonical expressions with structured metadata, such as whether the error is transient and a retry is advisable (e.g., for a rate-limiting error) or if it is a permanent failure (e.g., an invalid API key). This allows the consuming application to build robust, backend-agnostic error handling and resilience logic (e.g., retry mechanisms, circuit breakers) without needing to parse error strings or understand the failure modes of each individual database.

\section{Discussion and Future Work}

The development and evaluation of Vextra have demonstrated the viability and benefits of a unified abstraction layer for the heterogeneous vector database ecosystem. However, this work also highlights the inherent limitations and opens up promising avenues for further research.

\subsection{Limitations and Trade-offs of a Unified Abstraction}

The primary tradeoff of any middleware is performance overhead versus portability. Our evaluation in Section V-B quantified this, showing a low but nonzero latency increase. For applications with extremely stringent, sub millisecond latency requirements, direct native SDK integration may remain the preferred approach. 

Another significant challenge is the maintenance burden. The vector database landscape is evolving at a breakneck pace, with frequent updates to APIs and features. Maintaining the suite of Vextra adapters to ensure compatibility with the latest versions of each backend will require a continuous and dedicated effort. A breaking change in a database's API, for example, would require a corresponding update and release of its Vextra adapter to ensure uninterrupted service for users.

Finally, the `escape hatch' mechanism, while a pragmatic solution for accessing nonstandard features, represents a deliberate compromise in the abstraction. Over-reliance on this feature by an application will erode the benefits of portability. The long-term goal should be observe which provider-specific features gain widespread adoption and then promote them into the core, standardized Vextra API, minimizing the need for the escape hatch over time. 

\subsection{Opportunities for Advanced Query Optimization and Planning}
The introduction of a middleware layer like Vextra creates a strategic point of control and observation in the data path, enabling a new class of optimizations that are difficult or impossible to implement at the application level.

\begin{itemize}
    \item \textbf{Intelligent Query Caching}: Vextra is perfectly positioned to implement a query caching layer. It could cache the results of identical query calls (i.e., the same query vector and filter) for a configurable duration,  dramatically reducing latency and cost for applications with high-frequency, repetitive queries [39].
    \item \textbf{Automatic Request Batching}: Many applications may issue multiple small, concurrent queries. A more advanced version of Vextra could buffer these individual requests for a very short time window (e.g., 5-10 ms) and automatically batch them into a single, larger request to the backend database. Since most vector databases perform much more efficiently with batched operations, this could significantly increase overall throughput [40].
    \item \textbf{Cost-Based Query Optimizations in a Federated Context}: A future version of Vextra can be extended to manage multiple database backends simultaneously, acting as a federated query engine. In such a setup, if data is replicated or sharded across different types of databases (e.g., a high-performance, in-memory store and a low-cost disk based store), Vextra could implement a cost based optimizer. It could route queries to the most appropriate backend based on latency requirements, query complexity, and even the monetary cost associated with the query on different platforms [41]. 
\end{itemize}

\subsection{Towards a Self-Configuring and Extensible Middleware}
The current adapter model requires manual implementation of each new database. A long-term vision for Vextra is to evolve towards a more self-configuring and community-driven ecosystem.
\begin{itemize}
    \item \textbf{Dynamic Capability Discovery}: As vector databases mature, they may begin to expose standardized metadata endpoints that describe their capabilities, support filter operations, and indexing options (e.g., via an OpenAPI or Swagger specification). A future Vextra could query these endpoints to dynamically configure parts of its adapter, reducing the manual effort required to support a new database or a new version of an existing one [42].
    \item \textbf{Community-driven Adapter Framework}: To ensure Vextra can keep pace with the rapidly growing number of vector databases, we plan to formalize the adapter development process and provide a clear framework and tooling for community contributions. By creating a simple, well-documented process for developing, testing, and submitting new adapters, Vextra could leverage the collective effort of the open-source community to expand its support far more rapidly.
\end{itemize}

\section{Conclusion}

The proliferation of vector databases has been a pivotal enabler for the current wave of AI innovation, yet it has inadvertently created a fragmented and challenging development landscape. The lack of API standardization across these critical data systems imposes a significant tax on developers, stifling portability, increasing maintenance costs, and creating undesirable vendor lock-in.

In this paper, we have made three primary contributions to address this problem. First, we systematically identified and quantified the API fragmentation problem across the leading vector database solutions, providing concrete evidence of the challenges developers face. Second, we proposed, designed and implemented \textbf{Vextra}, a novel, middleware abstraction layer. Built on the proven Adapter Design pattern, Vextra provides a single, unified and intuitive API for all core vector database operations, effectively decoupling application logic from specific backend implementations. Third, we conducted a rigorous experimental evaluation that validated our approach. The results demonstrate that Vextra introduces minimal and acceptable performance overhead, typically in low single-digit percentages for production-style workloads, while offering a transformative improvement in application portability, as demonstrated by our seamless backend migration case study.

Vextra is more than just a software tool; it is a proposal for a path toward a more mature and interoperable vector database ecosystem. By providing a stable, common interface, it lowers the barrier to entry for developers, reduces the friction for adopting new database technologies, and creates a strategic point for future, higher level query optimizations. We conclude that a standardized abstraction is not merely a convenience but an essential piece of infrastructure required for continued growth, sustainability and democratization of AI application development. 

\section{AI-Generated Content Acknowledgement}
We have employed ChatGPT and Gemini, as research aids, to brainstorm literature review materials and to find potential keywords and references related to the topics explored in our paper. For instance, we used these tools as a gateway to search for database paradigms discussed previously and used it to inform our research with respect to the problem of API fragmentation. No AI technology was used to create diagrams, charts, code or text included in this paper.

\end{document}